\documentclass[preprint]{aastex631}

\usepackage{url} 
\usepackage{amsmath}
\usepackage{makecell}
\usepackage{comment}
\usepackage{natbib}

\submitjournal{APJ}

\begin{document}


\title{Ocean Tides on Asynchronously Rotating Planets Orbiting Low-mass Stars}

\author[0009-0008-8229-1994]{Jiaru Shi}
\affiliation{Laboratory for Climate and Ocean-Atmosphere Studies,
Department of Atmospheric and Oceanic Sciences, School of Physics,
Peking University, Beijing 100871, China}
\affiliation{Department of Earth, Atmosphere and Planetary Science, MIT, Cambridge, MA 02139, USA}

\author{Jun Yang}
\correspondingauthor{Jun Yang}
\affiliation{Laboratory for Climate and Ocean-Atmosphere Studies,
Department of Atmospheric and Oceanic Sciences, School of Physics,
Peking University, Beijing 100871, China}
\email{junyang@pku.edu.cn}

\author{Dorian S. Abbot}
\affiliation{Department of the Geophysical Sciences, University of Chicago, Chicago, IL 60637, USA}

\author{Yonggang Liu}
\affiliation{Laboratory for Climate and Ocean-Atmosphere Studies,
Department of Atmospheric and Oceanic Sciences, School of Physics,
Peking University, Beijing 100871, China}

\author{Wanying Kang}
\affiliation{Department of Earth, Atmosphere and Planetary Science, MIT, Cambridge, MA 02139, USA}

\author{Yufeng Lin}
\affiliation{Department of Earth and Space Sciences,
Southern University of Science and Technology, Shenzhen 518055, China}



\begin{abstract}

Planets in the liquid-water habitable zone of low-mass stars experience large tidal forces, $10^3$ to $10^4$ times those on Earth, due to the small distance between the habitable zone and the host stars. Therefore, interior solid tides, ocean tides and atmospheric tides on these planets could be much stronger than that on Earth, but rare work has been done to explicitly simulate the ocean tides. Here, for the first time, we perform global ocean tide simulations and show that ocean tides on asynchronously rotating planets with large eccentricities can reach $\mathcal{O}(1000)\,\mathrm{m}$ in height and $\mathcal{O}(10)\,\mathrm{m\,s^{-1}}$ in flow speed. Interactions between tide and bottom topography can induce large energy dissipation, $\sim\mathcal{O}(100)\,\mathrm{W\,m^{-2}}$ in global mean. This tidal energy dissipation can strongly accelerate orbital evolution by 1-2 orders of magnitude. However, for planets with small eccentricities, the ocean tides are much weaker but still comparable to that on modern Earth. Our results suggest that ocean tides on eccentric planets orbiting low-mass stars are orders of magnitude more powerful than those on Earth and can dramatically influence surface geography and orbital evolution.

\end{abstract}

\keywords{Exoplanet tides (497), Ocean planets (1151), Tidal distortion (1697), Tidal friction (1698)}


\section{Introduction} \label{sec:intro}

Ocean tides on a planet are oceanic wave-like motions that result from gravitational forces by its host star, satellite(s), or planetary companion(s). Ocean tides have important effects on planetary orbital evolution, the marine ecosystem, ocean mixing, and climate \citep{Webb80,Webb82,Stewart08,Green17,Motoyama20,Daher21,Green23}. Previous studies suggested that tidal dissipation rate on planets orbiting low-mass stars could be much greater than that on Earth and may be comparable to the stellar flux on certain planets \citep{Jackson08,Behounkova11,Barnes13,Driscoll15,Selsis13,Heller13,Barnes17,Barr18,Dobos19,Colose21}, but none of these studies explicitly considered ocean tides. Estimations based on highly idealized equilibrium tide theory \citep{Lingam18,Si22,Barnes23} implied that ocean tidal elevation on the planets could reach the order of 100 m. The equilibrium tide theory assumes that the system simultaneously and instantaneously achieves hydrostatic deformation under the tidal force. Ocean tides, however, are not in equilibrium and are more complex than those described by equilibrium tide theory \citep{Green23,Stewart08}. For instance, equilibrium tide theory can not accurately predict the strengths and spatial patterns of tidal elevation, current, and energy dissipation \citep{Green23}.

Explicit analysis and simulations of ocean tides on exoplanets have only recently begun to be made. \citet{Auclair18,Auclair19} focused on the dependence of oceanic tidal resonance on ocean depth, stratification and bottom drag, and the effect of ocean tides on planetary spin evolution, but the key characteristics of the tides such as tidal elevation and current speed were not shown in their studies. \citet{Blackledge20} simulated ocean tides and tidal dissipation under a series of random continental configurations but in present Earth’s orbit, so that their results cannot be directly applied to real exoplanets. To our knowledge, this work here is the first to explicitly simulate ocean tides on rocky exoplanets orbiting low-mass stars.\par


Rocky planets orbiting low-mass stars are easiest to detect, due to the large mass and size of the planets relative to the host stars, and because M and K dwarfs are the most abundant stellar type. Tidal force is inversely proportional to the cube of the star-planet distance. Planets in the liquid-water habitable zone of low-mass stars are therefore subjected to much larger tidal forces than those on Earth due to the small star-planet distance, less than 0.1 $\mathrm{AU}$, or even close to 0.01 $\mathrm{AU}$ \citep{Kasting93,Lyu24}. In this study, we employ a two-dimensional global shallow-water model over a sphere to simulate the ocean tides on asynchronously rotating planets orbiting M dwarf, with assuming that the planets have oceans. 
\par
We examine three exoplanets, Proxima Centauri b (labelled as ‘Proxima b’ hereafter), GJ 3323b, and TRAPPIST-1e. Their observed orbital eccentricities are $\leq 0.35$, $0.23 \pm 0.11$, and $0.005 \pm 0.0005$, respectively \citep{Anglada16,Brugger16,Astudillo17,Grimm18,Eric21}. Tidal potential is about 1778, 6947, and 3024 times Earth’s value, respectively (Table~\ref{tab:Tab1}). Observations shown that exoplanets cover a much wider range of orbital eccentricities (from 0 to 0.9) and have a much higher mean eccentricity than planets in the solar system, especially for single-transiting systems \citep{Limbach14,Sagear23}. Note that for synchronously rotating planets with zero eccentricity, tides would be steady, so they are not considered in this study.

\begin{table}[h!]
\caption{Parameters of Earth and the three potentially habitable exoplanets}
\centering
\begin{tabular}{l c c c c}
\toprule
Planets & Earth & Proxima b & GJ 3323b & TRAPPIST-1e \\
\hline
Star mass ($M_\odot$) & 1 & 0.12 & 0.17 & 0.08  \\
Planet mass ($M_\oplus$) & 1 & 1.27$^a$ & 2.02$^a$ & 0.62$^a$ \\
Planet radius ($a,\,\mathrm{km}$) & 6371 & 8282$^a$ & 7645$^a$ & 5861$^a$ \\
Planet gravity ($g,\,\mathrm{m\,s^{-2}}$) & 9.81 & 7.37$^a$ & 13.76$^a$ & 7.20$^a$ \\
Mean density ($\bar\rho,\,\mathrm{kg\,m^{-3}}$) & 5507 & 3183$^a$ & 6437$^a$ & 4413$^a$ \\
Orbital period (Earth days) & 365 & 11.19 & 5.36 & 6.10 \\
Rotation period (Earth days) & 1 & 7.46$^b$ & 3.57$^b$ & 6.10$^b$ \\
Semi-major axis ($r,\,\mathrm{AU}$) & 1 & 0.0485 & 0.0328 & 0.0282 \\
Global-mean stellar flux ($\mathrm{W\,m^{-2}}$) & 340 & 225 & 877 & 205 \\
Observed eccentricity & 0.0167 & $\leq0.35$ & $0.23\pm0.11$ & $0.0051\pm0.00058$ \\
Employed eccentricity & 0.0167 & 0.3$^a$ & 0.23$^a$ & 0.005$^a$ \\
Tidal potential ($\mathrm{0.80\,m^2\,s^{-2}}$)$^c$ & 1 & 1778 & 6947 & 3024 \\
Tidal acceleration ($\mathrm{2.5\times10^{-7}\,m\,s^{-2}}$)$^c$ & 1 & 2861 & 11164 & 4886 \\
\hline
\multicolumn{5}{l}{\parbox{16cm}{\textbf{Notes.}}}\\
\multicolumn{5}{l}{\parbox{16cm}{$^a$These parameters are not well known from observations.}}\\
\multicolumn{5}{l}{\parbox{16cm}{$^b$The rotation period is chosen following a 3:2 spin-orbit resonance state (like Mercury) for Proxima b and GJ 3323b, but 1:1 for TRAPPIST-1e due to its very small eccentricity.}}\\
\multicolumn{5}{l}{\parbox{16cm}{$^c$The values within parentheses are the corresponding tidal potential and tidal acceleration of solar semidiurnal tide on Earth.}}
\end{tabular}
\label{tab:Tab1}
\end{table}

In Section 2, we provide a comprehensive description of the methodology used in our simulations. Specifically, Section 2.1 outlines the planetary parameters of the exoplanets under investigation, while Section 2.2 details the governing equations of our ocean tidal model. In Section 2.3, we derive the tidal potential ($V_\mathrm{tid}$) at a given point on the planetary surface, and Section 2.4 describes the spatial configuration of land and ocean in our model.
Section 3 presents the results of our simulations. In Section 3.1, we analyze the tidal elevation ($\eta$) and tidal current ($\boldsymbol{u}$) across different planetary environments. Section 3.2 explores the sensitivity of $\eta$ and $\boldsymbol{u}$ to a broader range of key parameters, including the amplitude of the tidal potential, orbital eccentricity, ocean depth, and planetary rotation rate. Section 3.3 introduces the parameterization of tidal bottom drag in our model and examines the resulting tidal dissipation rate. Finally, in Section 3.4, we investigate the influence of oceanic tidal dissipation on the orbital evolution of exoplanets by performing numerical simulations using an orbital evolution model.

\section{Materials and Methods} \label{sec:method}
\subsection{Planetary Parameters}

The masses of the host stars of the three aforementioned planets applied in this study are 0.12 solar mass for Proxima b, 0.17 solar mass for GJ 3323b, and 0.08 solar mass for TRAPPIST-1e (Table~\ref{tab:Tab1}). The most recent studies have found the mass of TRAPPIST-1 to be 0.09$\pm0.001$ solar mass \citep{Birky21,Agol21}, around 10\% higher than our applied value, which won't lead to a qualitative change in our result. Despite the relatively low mass of their host stars, their tidal forces are much stronger than Earth's. This is primarily due to their closer distances to the host stars, which are 0.0485 $\mathrm{AU}$ for Proxima b, 0.0328 $\mathrm{AU}$ for GJ 3323b, and 0.0282 $\mathrm{AU}$ for TRAPPIST-1e. Another important parameter to determine tidal force is the planetary radius, which is relatively well constrained given the solid nature of these planets. The exact planetary radius depends on the planet's composition and internal structure, leading to an uncertainty of no more than 50\% in the exact radius estimates for the planets considered in this study \citep{Brugger16,Eric21}. Based on current observational constraints, we adopt planetary radii of 1.3 times Earth's radius for Proxima b, 1.2 times Earth's radius for GJ 3323b, and 0.92 times Earth's radius for TRAPPIST-1e.
\par
The exact eccentricities of the planets are unknown, but their upper limits or ranges can be determined from observations (Table~\ref{tab:Tab1}). By default, eccentricity values used are 0.30 for Proxima b, 0.23 for GJ 3323b, and 0.005 for TRAPPIST-1e. The exoplanets' rotation periods are unobserved, so we assume moderate values: 2/3 of the orbital periods for Proxima b and GJ 3323b (like Mercury) due to their relatively high eccentricity, and equal to the orbital period for TRAPPIST-1e. Planetary obliquity is set to zero, as it should damp within $\sim10^7$ Earth years (e.g., \citet{Heller11}). Sensitivity experiments testing tidal potential, orbital eccentricity, ocean depth, and rotation period are detailed in Section 3.

\subsection{Ocean Tidal Model}
The ocean tidal model we used is a nonlinear shallow-water model with constant seawater density, based on the framework of MITgcm \citep{Marshall97,Campin04}. The main equations are as follows:
\begin{linenomath*}
    \begin{equation}
        \frac{D\boldsymbol u}{Dt}+2(\boldsymbol{u\times} \boldsymbol{f_c})=-g\boldsymbol{\nabla_h}\eta-\boldsymbol{\nabla_h} V_{tid}+A_h\mathit{\nabla_h}^2\boldsymbol u-\boldsymbol F,
        \label{eq:eq1}
    \end{equation}
    \begin{equation}
        \frac{\partial\eta}{\partial t}+\boldsymbol{\nabla_h\cdot}\int_{-H+\eta_b}^\eta \boldsymbol u dz=0,
        \label{eq:eq2}
    \end{equation}
    \begin{equation}
        \boldsymbol F = \mathbf{C_\mathrm{tid}}\boldsymbol{\cdot u}/H + C_D |\boldsymbol u|\boldsymbol u/H,
        \label{eq:eq3}
    \end{equation}
\end{linenomath*}
where $\boldsymbol{u}$ is the horizontal tidal current velocity, $\eta$ is the sea surface elevation relative to sea level, $\boldsymbol{f_c}$ is the Coriolis parameter, $g$ is the planetary gravity, $V_\mathrm{tid}$ is the tidal potential, $A_h$ is the horizontal eddy viscosity, $\boldsymbol{F}$ is the frictional or dissipative stress term, the subscript $h$ indicates horizontal direction, $\eta_b$ is the ocean bottom height relative to the mean ocean depth ($-H$), $C_D$ is the bottom drag coefficient, and $\mathbf{C_\mathrm{tid}}$ is an energy conversion coefficient with units of $\mathrm{m\,s^{-1}}$. The friction term ($\boldsymbol{F}$) includes two parts: energy loss due to tidal conversion from large scale, vertically homogeneous tidal currents (barotropic tide) to small scale, three-dimensional, vertically inhomogeneous internal gravity waves (baroclinic tide) over rough bathymetry (such as mid-oceanic ridges), such as mid-oceanic ridges, and quadratic drag due to ocean bottom stress in the frictional boundary layer of shallow seas \citep{Mofjeld88}. The value of $C_D$ is a constant, 0.003. Previous sensitivity tests showed that simulation results are insensitive to the value of $C_D$ as long as it is in the range of 0.001 to 0.01 \citep{Egbert04}. $\mathbf{C_\mathrm{tid}}$ is a 2×2 tensor, we will derive it in section 3.3. By default, the model time step is 100 s, small enough to keep the system stable under the Courant–Friedrichs–Lewy (CFL) condition. For sensitivity tests of ocean depth and rotational rate, the time step is decreased accordingly due to stronger ocean currents. 

\subsection{Tidal Potential}
The tidal potential ($V_\mathrm{tid}$) includes two parts, the astronomical forcing from the host star adjusted by the deformation of the solid planet ($V_\mathrm{ex}$) and the perturbation potential caused by the mass redistribution of ocean ($V_\mathrm{load}$), following the methods of \citet{Ray98} and \citet{Agnew07}:
\begin{linenomath*}
    \begin{equation}
        V_\mathrm{tid}=V_\mathrm{ex}+V_\mathrm{load},
        \label{eq:eq4}
    \end{equation}
    \begin{equation}
        V_\mathrm{ex}=-\frac{GM}{d}=-\frac{GM}{r}\sum_{n=0}^\infty \Big(\frac{a}{r}\Big)^n\frac{4\pi\gamma_n}{2n+1}\sum_{m=-n}^n Y_n^m(\lambda,\phi)Y_n^{m*}(\frac{\pi}{2},\delta),
        \label{eq:eq5}
    \end{equation}
    \begin{equation}
        V_\mathrm{load}\approx-\frac{Gm_p}{a}\frac{\rho_w}{\bar\rho}\sum_{n=0}^\infty\frac{3\gamma_n'}{2n+1}\sum_{m=-n}^{n}\frac{\eta_n^m}{a}Y_n^m(\lambda,\phi),
        \label{eq:eq6}
    \end{equation}
\end{linenomath*}
where $G$ is the gravitational constant, $a$ is the planet’s average radius, $M$ is the star’s mass, $d$ is the distance between observational point and the star, $r$ is the orbital distance between the planet's center and the star's center, $Y_n^m$ is the spherical harmonic function of degree $n$ and order $m$, $\lambda$ is colatitude, $\phi$ is longitude, $\delta$ is the true anomaly, $k_2$ and $h_2$ are Love numbers, $m_p$ is the planet’s mass, $\rho_w$ is seawater density ($\mathrm{\sim1000\,kg\,m^{-3}}$), $\bar\rho$ is the planet’s mean density, and $\eta_n^m$ is the spherical harmonics coefficient of degree $n$ and order $m$ for the sea surface elevation ($\eta$). $\gamma_n$ is a ‘diminishing factor’, which represents the effects of self-attraction by the deformation of the solid planet and the vertical displacement of the solid surface \citep{Love09,Love11,Efroimsky12,Correia14}.\par

For all the experiments, we focus on the principal component of the stellar tides: $n=2$, $\gamma_n=\gamma_2$. According to \citet{Love11}:
\begin{linenomath*}
    \begin{equation}
        \gamma_2=1+k_2-h_2,\quad h_2=\frac{5}{2}\Big(1+\frac{19}{2}\frac{\mu}{g\bar\rho a}\Big)^{-1},\quad k_2=\frac{3}{5}h_2,
        \label{eq:eq7}
    \end{equation}
\end{linenomath*}
where $k_2$ and $h_2$ are Love numbers, $k_2$ indicates the self-attraction deformation, $h_2$ indicates the vertical deformation, $\mu$ is the average rigidity of the planet, and $g$ is the planet’s surface gravity. Similarly, $\gamma_n'$ represents the effects of self-attraction and vertical displacement of the solid body caused by the ocean’s perturbation potential. If we only consider each mass point’s loading on its neighboring region, we can obtain the following expression for $\gamma_n'$:
\begin{linenomath*}
    \begin{equation}
        \gamma_n'=1+k_n'-h_n',\quad k_n'=g^2k_{n,\mathrm{E}}'/g_\mathrm{E}^2,\quad h_n'=g^2h_{n,\mathrm{E}}'/g_\mathrm{E}^2,
        \label{eq:eq8}
    \end{equation}
\end{linenomath*}
where $k_n'$ and $h_n'$ are the loading Love numbers, $k_{n,\mathrm{E}}'$ and $h_{n,\mathrm{E}}'$ are the corresponding values for Earth ($-$0.31 and $-$1.00, respectively \citep{Ferrall72}), and $g_\mathrm{E}$ is the surface gravity of Earth, with assuming the Lam\'{e} parameters are the same as Earth’s.

Equation~(\ref{eq:eq6}) shows that $V_\mathrm{load}$ depends on the tidal elevation ($\eta$), which makes it hard (if not impossible) to calculate this term synchronously. However, $\eta_2^2$ is much bigger than other components, so $V_\mathrm{load}$ can be simplified as \citep{Ray98,Ferrall72}:
\begin{linenomath*}
    \begin{equation}
        V_\mathrm{load}\approx \frac{3\gamma_2'\rho_2/(5\bar\rho)}{1+3\gamma_2'\rho_w/(5\bar\rho)}V_\mathrm{ex}\equiv\sigma V_\mathrm{ex}.
        \label{eq:eq9}
    \end{equation}
\end{linenomath*}
Based on the parameters shown in Table~\ref{tab:Tab1} and the value of $\mu$ ($\mathrm{\sim1.13\times10^{11}\,kg\,m^{-1}\,s^{-2}}$, assuming it to be equal to Earth’s value \citep{Balakrishna67}), we can obtain the Love numbers and $\sigma$. The results are summarized in Table~\ref{tab:Tab4}. The uncertainties in these numbers should not be small. However, they should not influence our main conclusions, based on the fact that the tidal potential from the host stars on the planets is several orders larger than that on Earth, as discussed in \citet{Pierrehumbert19}.\par

\begin{table}[h!]
    \centering
    \hskip-2.0cm
    \begin{tabular}{l c c c c c c c}
        \toprule
        Planets & $h_2$ & $k_2$ & $\gamma_2$ & $h_2'$ & $k_2'$ & $\gamma_2'$ & $\sigma$ \\
        \hline
        Earth & 0.61 & 0.36 & 0.75 & -1.00 & -0.31 & 1.69 & 0.20 \\
        Proxima b & 0.38 & 0.23 & 0.85 & -0.56 & -0.18 & 1.38 & 0.20 \\
        GJ 3323b & 0.97 & 0.58 & 0.61 & -1.97 & -0.61 & 2.36 & 0.18 \\
        TRAPPIST-1e & 0.37 & 0.22 & 0.85 & -0.54 & -0.17 & 1.37 & 0.16 \\
        \hline
    \end{tabular}
    \caption{Love numbers and loading Love numbers for Earth and the three exoplanets in this study. The definitions of the parameters are elaborated by Equations~(\ref{eq:eq7}), (\ref{eq:eq8}), and (\ref{eq:eq9}).}
    \label{tab:Tab4}
\end{table}

\begin{figure}[h!]
\noindent\includegraphics[width=\textwidth]{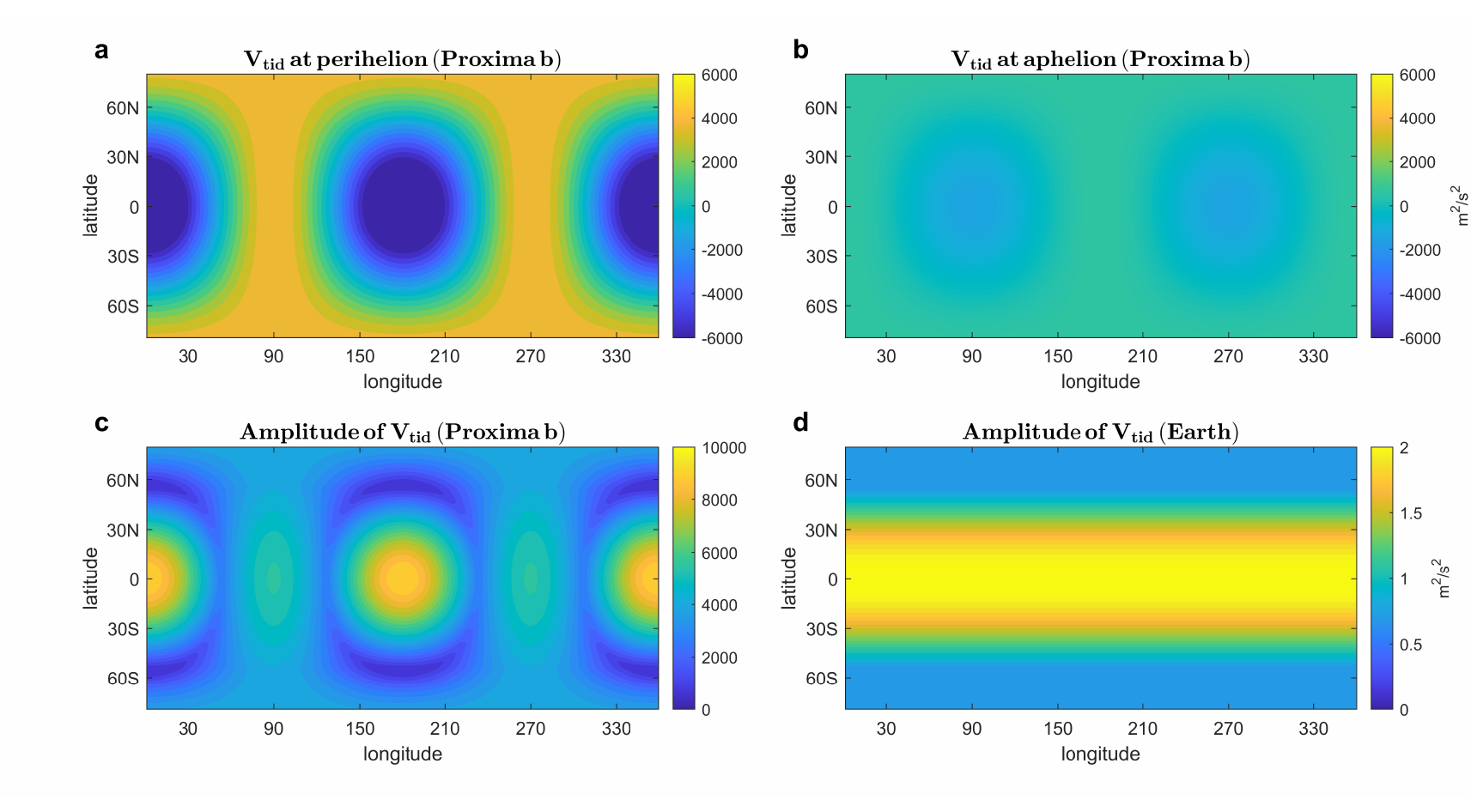}
\centering
\caption{The instantaneous $V_\mathrm{tid}$ on Proxima b at periastron (a) and at apastron (b), and the orbital-mean amplitude of tidal potential ($V_\mathrm{tid}$) on Proxima b (c) and on Earth (d).}
\label{fig:Fig1}
\end{figure}

Simultaneously solving Equations~(\ref{eq:eq4}) (\ref{eq:eq5}) (\ref{eq:eq6}) and (\ref{eq:eq9}), we obtain $V_\mathrm{tid}$ for each planet and visualize the spatial distribution of $V_\mathrm{tid}$ on Proxima b at its periastron and apastron in Figure~\ref{fig:Fig1}a \& b. The spatial distribution of tidal potential on Proxima b is similar to that on Earth induced by the sun. The global maximum of tidal potential reaches $\mathcal{O}(10^4)\,\mathrm{m^2\,s^{-2}}$ at the periastron, about one order of magnitude larger than the global maximum of tidal potential at the apastron. This variation in $V_\mathrm{tid}$ at periastron and apastron is attributed to the high eccentricity of Proxima b. According to Equation~(\ref{eq:eq5}), $V_\mathrm{tid}$ is proportional to $r^{-3}$, while $r$ varies by over 50\% of its mean value during a single tidal period due to the planet's high eccentricity. Figure~\ref{fig:Fig1}c \& d depict the time averaged amplitude of tidal potential, defined as the difference between its maximum and minimum values over one tidal period at a given spatial point. When comparing Proxima b to Earth, the tidal amplitude on Proxima b is about 4 orders of magnitude higher. Additionally, a longitudinal asymmetry emerges in the spatial distribution of tidal amplitude on Proxima b, which is attributed to the 3:2 spin-orbit resonance of Proxima b. This phenomenon will be further discussed in Section 3.3.

\subsection{Ocean Bathymetry}
Two different land-ocean configurations are examined. One is modern Earth’s land-ocean configuration, and the other one is an aqua-planet without any continent. For the Earth’s configuration experiment, the ocean depth is set to be 3688 m. For the aqua-planet experiments, ocean depth is set to be 10,000 m. For both Earth’s configuration and aqua-planet configuration, the spatial deviation of the ocean depth is set to be the mean value of Earth’s oceans, 1325 m, and the horizontal gradient of the ocean bottom depth ($|\boldsymbol{\nabla_h} \eta_b|$) is also set to be the mean value of Earth’s oceans, 0.018. In the ocean tide experiments, the grid spacing is $0.75^\circ$ by latitude and $0.8^\circ$ by longitude over a nearly global sphere, covering the area from $80^\circ$S to $80^\circ$N. Polar grids are not allowed in the framework of MITgcm \citep{Marshall97}.

\section{Results} \label{sec:result}
\subsection{Tidal Elevation and Tidal Currents}


\begin{figure}[h!]
\noindent\includegraphics[width=\textwidth]{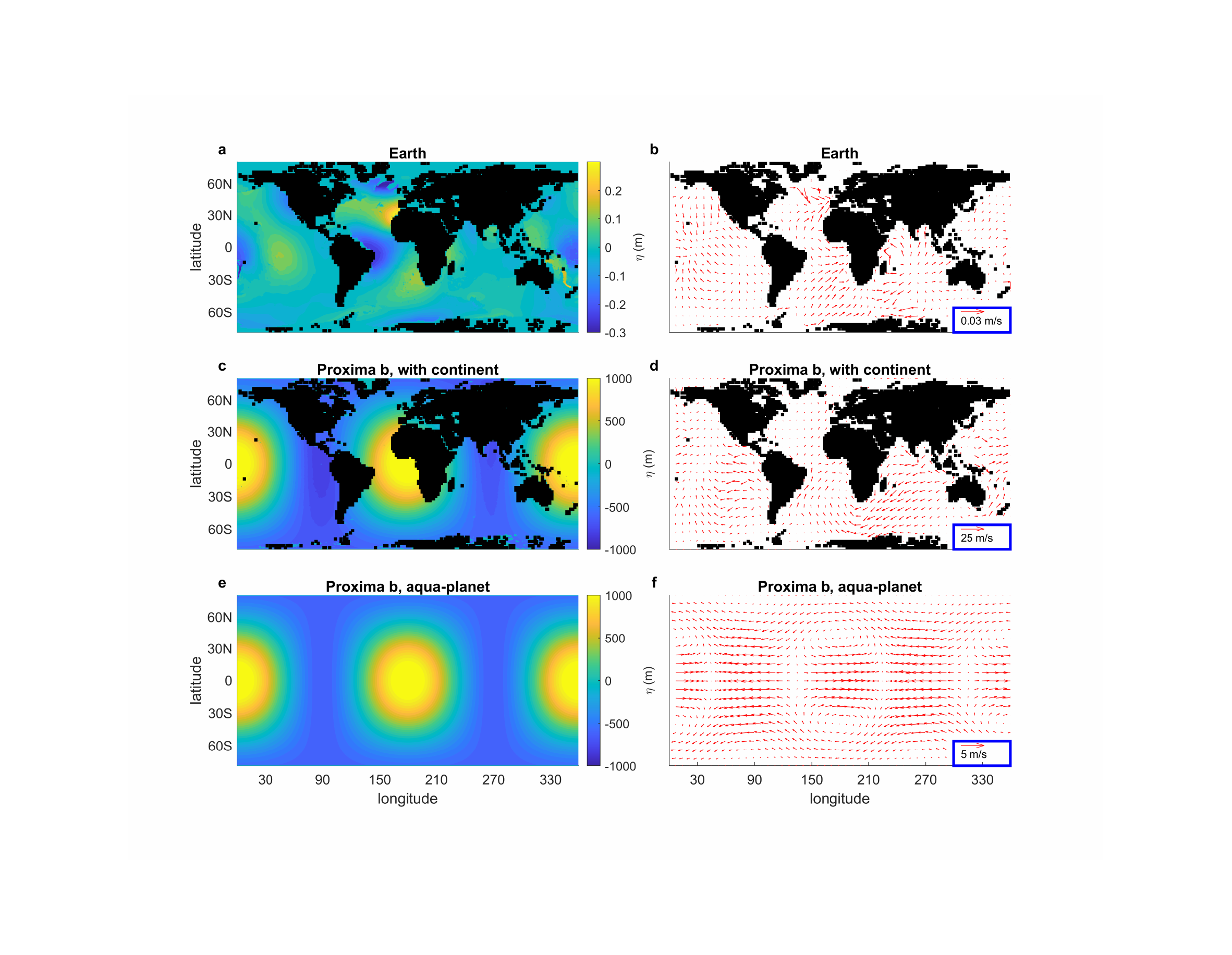}
\centering
\caption{Snapshots of simulated tidal elevation (left) and tidal currents (right) on Earth and Proxima b. Land-ocean configurations of modern Earth (a-d) and an aqua-planet without any continent (e-f) are used.}
\label{fig:Fig2}
\end{figure}

\begin{figure}[h!]
\noindent\includegraphics[width=\textwidth]{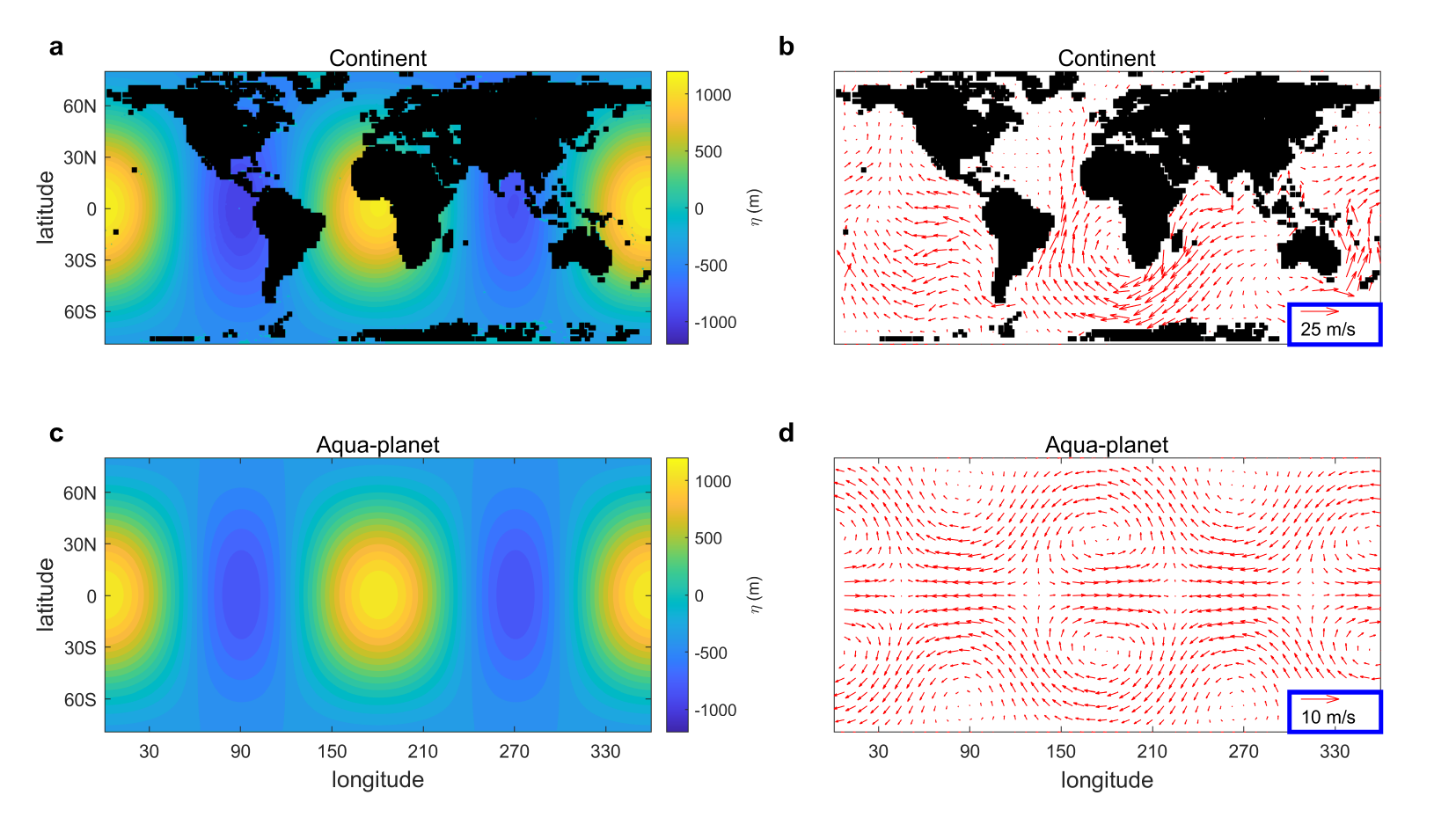}
\centering
\caption{Snapshots of simulated tidal elevation and tidal current on GJ 3323b. Upper panels are for the experiments with modern Earth’s land-ocean configuration, and lower panels are for the experiments with no continent (i.e., aqua-planet). Tidal elevation and tidal current on GJ 3323b are 2-3 orders stronger than those on Earth.}
\label{fig:Fig3}
\end{figure}

For Earth, the simulated tidal elevation is in the order of $\mathrm{0.1\mbox{-}1.0\,m}$ (Figure~\ref{fig:Fig2}a) and the global root mean square (RMS) of the tidal elevation is 0.09 m. These results are in the range of observations and are consistent with previous simulations \citep{Egbert01,Egbert03,Schindelegger18,Lyard21}, suggesting that the model we use provides a reasonable approximation of Earth’s ocean tides. For Proxima b, the tidal elevation is in the order of 1000 m and its global-mean RMS is $\sim$200 m, regardless of the employed land-ocean configuration (modern Earth or aqua-planet), as shown in Figure~\ref{fig:Fig2}c \& e. For GJ 3323b, the simulated tidal elevation can reach $\sim$1200 m and the global RMS is $\sim$300 m (Figure~\ref{fig:Fig3}a \& c). Our simulation results are about one order of magnitude larger than the estimations based on simple equilibrium tide theory \citep{Si22,Barnes23,Lingam18}.\par

The tidal currents on Proxima b are in the order of $\mathrm{10\,m\,s^{-1}}$ and the strongest current can reach $\mathrm{25\,m\,s^{-1}}$, 2-3 orders stronger than that on Earth (Figure~\ref{fig:Fig2}b, d \& f). Under the continental configuration of modern Earth, tidal currents are several times stronger than those on an aqua-planet, in particularly in coastal regions (Figure~\ref{fig:Fig2}d).

For GJ 3323b, tidal currents are even stronger (Figure~\ref{fig:Fig3}b \& d). The strength of the tide-forced ocean currents on both Proxima b and GJ 3323b is several to ten times that of wind-driven ocean currents \citep{Hu14,Del19,Yang19}.\par

\begin{table}[h!]
    \caption{Scale analysis of the momentum equation for the control experiment of Proxima b under an aqua-planet configuration}
    \centering
    \hskip-1cm
    \begin{tabular}{l c c c}
        \toprule
        Terms & Expression & Scaling & \makecell[l]{Magnitude\\($\mathrm{m\,s^{-2}}$)} \\
        \hline
        Tidal acceleration & $\boldsymbol{\nabla_h}V_\mathrm{tid}$ & $\frac{GMa}{R^3}$ & 10E-4 \\
        Pressure gradient & $g\boldsymbol{\nabla_h}\eta$ & $\frac{g\eta^*}{a}$ & 10E-4 \\
        Coriolis force & $2(\boldsymbol{\Omega_h\times u})$ & $\Omega U^*$ & 10E-5 \\
        Inertial term (horizontal) & $\partial_t\boldsymbol{u}$ & $\Omega U^*$ & 10E-5 \\
        Tidal bottom drag & $\frac{\mathbf{C}_\mathrm{tid}\boldsymbol{\cdot u}}{H}$ & $\frac{C_\mathrm{tid}}{H}U^*$ & 10E-5 \\
        Quadratic drag & $\frac{C_D}{H}|\boldsymbol{u}|\boldsymbol{u}$ & $\frac{C_D}{H}U^{*2}$ & 10E-7 \\
        Advection term & $(\boldsymbol{u\cdot\nabla})\boldsymbol{u}$ & $\frac{U^{*2}}{a}$ & 10E-7 \\
        Inertial term (vertical) & $\partial_tw$ & $\Omega\frac{H}{a} U^*$ & 10E-8 \\
        Viscosity term & $A_h\nabla_h^2\boldsymbol{u}$ & $\frac{A_hU^*}{a^2}$ & 10E-9 \\
        \hline
    \end{tabular}
    \label{tab:Tab2}
\end{table}

A scale analysis for different forces on Proxima b and GJ 3323b is summerized in Table~\ref{tab:Tab2}. In scaling the different terms, $V_\mathrm{tid}$ is in the order of $10^3\,\mathrm{m^2\,s^{-2}}$, $M$ is the star’s mass, $R$ is the star-planet distance, $U^*$ is the typical value of the tidal currents ($\boldsymbol{u}$, in the order of $1.0\,\mathrm{m\,s^{-1}}$), $\eta^*$ is the typical value of tidal elevation ($\eta$, in the order of $10^2\,\mathrm{m}$), $g$ is the planet’s gravity (in the order of $10\,\mathrm{m\,s^{-2}}$), $\Omega$ is the planetary rotation rate (in the order of $10^{-5}\,\mathrm{s^{-1}}$), $H$ is the ocean depth (in the order of $10^4\,\mathrm{m}$ for the open ocean), $a$ is the planet’s radius (also being equal to the typical horizontal length scale, in the order of $10^7\,\mathrm{m}$), $C_\mathrm{tid}$ is the tidal conversion coefficient (in the order of $0.1\,\mathrm{m\,s^{-1}}$), $C_D$ is the bottom drag coefficient in shallow oceans (in the order of $10^{-3}$), and $A_h$ is the horizontal eddy viscosity (in the order of $10^5\,\mathrm{m^2\,s^{-1}}$). 
The scale analysis reveals that the dominant force balance governing the system is between the pressure gradient force and the tidal force. Secondary contributions from the Coriolis force, inertial effects, and tidal bottom drag influence the phase lag between the tidal force and the tidal bulge. As tidal currents increase in magnitude, these secondary forces also intensify, leading to a greater phase lag. The nonlinear terms, including quadratic bottom drag and advection, are comparatively negligible. Consequently, tidal elevation and tidal currents remain approximately proportional to the tidal force, a relationship that will be further examined through sensitivity experiments in Section 3.2.

\begin{figure}[h!]
\noindent\includegraphics[width=1\textwidth]{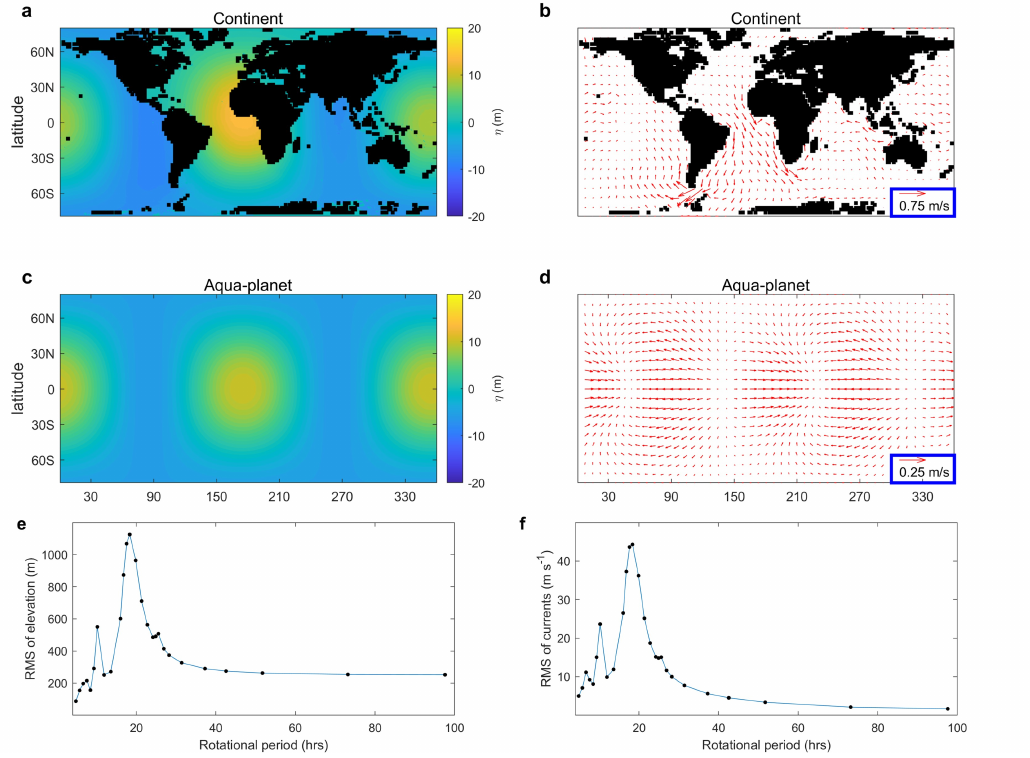}
\centering
\caption{Snapshots of simulated tidal elevation, tidal currents and their dependence on rotational rate on TRAPPIST-1e. Orbital eccentricity is set to be 0.005 in (a-d) and zero in (e-f). Tidal elevation and tidal currents on TRAPPIST-1e are weaker than those on Proxima b and GJ 3323b, but still larger than those on Earth. In (e-f), planetary parameters of TRAPPIST-1e are employed except that the rotation period is varied. Resonance occurs at a rotation period of $\sim$19 hrs, and a second peak exists at around $\sim$9.5 hrs.}
\label{fig:Fig4}
\end{figure}

The orbital eccentricity of TRAPPIST-1e ($<0.01$) is small and the age of the TRAPPIST-1 system is $7.6 \pm 2.2$ billion years \citep{Burgasser19}. So, TRAPPIST-1e may have already evolved to be exactly in or close to a synchronously rotating state \citep{Renaud20}. Here, we consider an eccentricity of 0.005 (close to the value of Jupiter’s moon Io, 0.0041), and assume that rotation period is equal to orbital period. The non-zero eccentricity has two effects: (1) the tidal force at periastron is larger than that at apastron; and (2) orbital angular velocity is a function of time, so the substellar point librates by $\sim$120 km. Both of these effects can cause ocean tides. The simulated tidal elevation is in the order of 10 m and the current speed is in the order of $\mathrm{0.1\,m\,s^{-1}}$ (Figure~\ref{fig:Fig4}a-d), both of which are about two orders of magnitude smaller than those on Proxima b and GJ 3323b but still larger than those on Earth. These results suggest that for planets with small eccentricities orbiting low-mass stars, ocean tides will not be strong but still non-negligible.

Spatial patterns of instantaneous tidal elevation and currents on Proxima b, TRAPPIST-1e, and GJ 3323b have a clear wavenumber-2 pattern with a pair of bulges, which is much clearer than that on Earth (Figures~\ref{fig:Fig2}, \ref{fig:Fig3} and \ref{fig:Fig4}). This is due to the slower rotation rate $\Omega$ (1/7.46 Earth’s for Proxima b, 1/6.1 Earth's for TRAPPIST-1e, and 1/3.57 Earth’s for GJ 3323b) on those exoplanets than on Earth, which leads to smaller tidal currents $U^*$ (see its estimation above). As a result, according to Table~\ref{tab:Tab2}, the inertial term, Coriolis term, and advection term will contribute less to the balance between major forces, surface elevation $\eta$ therefore matches better to the instantaneous tidal potential. In other words, slower rotation rate means longer tidal period, while longer tidal period means the ocean has more time to adjust to match the spatial pattern of tidal force.

\begin{table}[h!]
\caption{Main results of the control experiments and sensitivity tests}
\centering
\scriptsize
\begin{tabular}{l l c c c c c}
\toprule
Groups$^a$ & Experiments$^b$ & \makecell{$\eta_{\text{\tiny $\mathrm{max}$}}$\\(m)} & \makecell{$RMS_{\text{\tiny $\eta$}}$\\(m)} & \makecell{$RMS_{\text{\tiny $U$}}$\\($\mathrm{m\,s^{-1}}$)} & \makecell{$\bar D$\\($\mathrm{W\,m^{-2}}$)} & $Q$\\
\hline
Earth & Control experiment & 0.4 & 0.07 & 0.004 & 0.002 & 10 \\
\hline
Proxima b  & Control, with continents & 1081 & 216 & 2.6 & 192 & 5 \\
& Control, aqua-planet & 978 & 206 & 0.8 & 17 & 53 \\
\hline
GJ 3323b  & Control, with continents & 1188 & 299 & 5.4 & 1361 & 74 \\
& Control, aqua-planet & 1158 & 285 & 2.1 & 58 & 1726 \\
\hline
TRAPPIST-1e  & Control, with continents & 14 & 5 & 0.08 & 0.2 & 8 \\
& Control, aqua-planet & 12 & 5 & 0.05 & 0.02 & 84 \\
\hline
TRAPPIST-1e  & Rotation period=98 hrs & 492 & 252 & 1.7 & 65 & 135 \\
& Rotation period=52 hrs & 525 & 262 & 3.4 & 857 & 42 \\
& $^c$Rotation period=18.5 hrs & 3896 & 1125 & 44.1 & 5.7E5 & 0.2 \\
& $^c$Rotation period=9 hrs & 727 & 291 & 15.0 & 4.5E4 & 6 \\
\hline
Proxima b & Ocean depth=40 km & 719 & 154 & 0.1 & 0.5 & 1914 \\
& Ocean depth=20 km & 723 & 155 & 0.3 & 2.1 & 456 \\
& Ocean depth=5 km & 1017 & 208 & 1.6 & 104 & 9 \\
& Ocean depth=2.5 km & 1028 & 214 & 3.3 & 634 & 2 \\
\hline
Proxima b & Tidal potential $\times$ 0.3 & 340 & 72 & 0.3 & 2.0 & 53 \\
& Tidal potential $\times$ 0.1 & 108 & 23 & 0.09 & 0.2 & 53 \\
& Tidal potential $\times$ 0.01 & 11 & 2 & 0.009 & 0.002 & 53 \\
\hline
Proxima b  & Eccentricity=0.20 & 675 & 183 & 0.7 & 11 & 105 \\
& Eccentricity=0.25 & 847 & 202 & 0.8 & 14 & 76 \\
& Eccentricity=0.35 & 1397 & 262 & 1.2 & 58 & 14 \\
& Eccentricity=0.40 & 1855 & 316 & 2.3 & 282 & 2 \\
\hline
\multicolumn{7}{l}{\parbox{17cm}{\textbf{Notes.}}}\\
\multicolumn{7}{l}{\parbox{17cm}{$^a$Four groups are examined in the sensitivity experiments, including tidal potential, orbital eccentricity, ocean depth, and rotation period. For simplicity, all the sensitivity experiments employ the aqua-planet configuration and the default parameters of Proxima b or TRAPPIST-1e except if noted.}}\\
\multicolumn{7}{l}{\parbox{17cm}{$^b$The five variables listed from left to right are the maximum value of tidal elevation ($\eta_\mathrm{max}$), the global root mean squares of tidal elevation ($RMS_\eta$) and of tidal current speed ($RMS_U$), global mean tidal dissipation ($\bar D$), and tidal quality factor ($Q$).}}\\
\multicolumn{7}{l}{\parbox{17cm}{$^c$Tidal resonance happens within the ocean in these experiments.}}\\
\end{tabular}
\label{tab:Tab3}
\end{table}

\subsection{Sensitivity Tests}
Several key parameters are not constrained by observations, such as tidal potential, rotation period, the exact value of eccentricity, and ocean depth. To investigate the influences of these uncertainties, we did four groups of sensitivity experiments. The tidal elevation, tidal currents, and tidal dissipation become smaller if the tidal potential is weaker, orbital eccentricity is lower, or the ocean is deeper (Table~\ref{tab:Tab3}).\par

When varying rotation period, tidal elevation and current speed are non-monotonic functions, and resonance occurs when the rotation period is equal to $\sim$19 hrs for TRAPPIST-1e with aqua-planet configuration (Figure~\ref{fig:Fig4}e \& f). In the resonant state, the maximum tidal elevation reaches $\sim$3896 m, the RMS of tidal elevation is $\sim$1125 m, the RMS of tidal current speed is as high as $\sim$$\mathrm{44\,m\,s^{-1}}$. The resonance occurs when the frequency of tidal force is close to that of shallow-water gravity waves (free oscillation of the global ocean). The dominant tidal wavelength along the equator is $\pi a$. The phase speed of surface gravity waves can be estimated by $\sqrt{gH}$. Therefore, the estimated resonance period is $\pi a/\sqrt{gH}$, which is equal to 19.1 hrs for TRAPPIST-1e. Similar resonance phenomena have been found in the studies of stars \citep{Zahn70,Zahn75,Ogilvie14}, icy moons \citep{Tyler08,Chen14,Matsuyama14,Matsuyama18,Beuthe16}, Earth \citep{Chapman1924,Chapman70,Webb80,Zahnle87,Wei21,Deitrick24}, and other planets \citep{Zurek76,Ogilvie09,Ogilvie13,Deitrick24}.

The effect of varying ocean depth can be understood through the scaling addressed above, $U^*\approx a\Omega\eta^*/H$. When the ocean depth is changed, the mean tidal elevation ($\eta^*$) is almost unchanged; this is because $\eta^*$ is mainly constrained by the balance between tidal force and pressure gradient force, both of which are nearly independent of the ocean depth. So, when the ocean depth increases, the denominator of the scaling enhances and thereby the tidal current speed decreases. For instance, keeping $\eta^*$ unchanged, when the ocean depth is increased from 2.5 to 40 km, tidal current velocity decreases from $\sim$3.3 to $\mathrm{0.1\,m\,s^{-1}}$ (Table~\ref{tab:Tab3}).\par


Another poorly constrained parameter is orbital eccentricity, which shows a positive correlation with tidal elevation and tidal current in Table~\ref{tab:Tab3}. With nonzero $e$, the planet gets closer to the star when approaching periastron, and the closest distance decreases with increasing $e$. This ensures the planet a larger maximum of tidal potential, inducing a larger $\eta_\mathrm{max}$. This also ensures the planet a larger standard deviation in tidal elevation and tidal current while keeping their orbital averages unchanged, implying larger root mean square values of $\eta$ and $\boldsymbol{u}$.

\subsection{Tidal Drag and Dissipation Processes}
Interactions between tidal currents and seafloor topography can trigger surface and internal waves/eddies and small-scale turbulence, causing tidal energy to dissipate in shallow and deep oceans. The tidal dissipation processes are complex and there are large uncertainties, which is why the dissipation processes are usually parameterized by a horizontal drag at the bottom boundary in large scale simulation. Here, we discuss whether or not the tidal conversion drag parameterization that applied to Earth could be applied to exoplanets, and introduce a modification to it. According to \citet{Egbert04}, the tidal drag tensor (Equation~{\ref{eq:eq3}}) may be written as:
\begin{equation}
    \mathbf{C}_\mathrm{tid}=\frac{\langle P\rangle}{\rho_w|\boldsymbol{u\cdot\nabla}\eta_b|^2}(\boldsymbol{\nabla_h}\eta_b)^T\boldsymbol{\nabla_h}\eta_b,
    \label{eq:eq10}
\end{equation}
where $\langle P\rangle$ is the time average power of tidal conversion drag per area, $\langle\cdot\rangle$ denotes time average, and $\rho_w$ is the mean seawater density. For 1-D flows, $\langle P\rangle$ is estimated by:
\begin{equation}
    \langle P\rangle\approx\frac{\rho_wN_b\eta_b^2u^2}{4\sqrt{\pi}X_0}(1-f^2/\omega^2_0)^{0.5},
    \label{eq:eq11}
\end{equation}
where $N_b$ is the buoyancy frequency near ocean bottom, $X_0$ is the horizontal scale of bottom topography, $f$ is the Coriolis parameter, and $\omega_0$ is tidal frequency \citep{Llewellyn02}. We set $X_0$ to 100 km for all planets. Equation~(\ref{eq:eq11}) requires that $U_0\ll\omega_0 X_0$, where $U_0$ is the amplitude of tidal current velocity \citep{Bell75a}. This condition is well satisfied on Earth, but for the planets employed in this study, the tidal current is much faster and the tidal frequency is relatively low, which leads to $U_0\sim \omega_0X_0$ or even $U_0\gg\omega_0X_0$. Higher order modes ($n>1$) are induced and contribute to tidal conversion. 
\par
To be more general in the simulations of exoplanets, we applied a more accurate form of $\langle P\rangle$ \citep{Bell75b} to parameterize tidal conversion drag: 

\begin{equation}
    \langle P\rangle = \langle\overline{p\boldsymbol{u\cdot\nabla}\eta_b}\rangle\approx\frac{\rho_w\omega_0^2N_b}{2\pi^2X_0^2}\iint_{-\infty}^\infty\frac{dkdl}{\sqrt{k^2+l^2}}|\hat{\eta_b}|^2\sum_{n=n_0'}^{n=n_0}n^2\sqrt{(1-\frac{n^2\omega_0^2}{N_b^2})(1-\frac{f^2}{n^2\omega_0^2})}J_n^2(\beta),
    \label{eq:eq12}
\end{equation}
where $p$ is baroclinic bottom pressure, $N_b$ is the buoyancy frequency near ocean bottom, $n_0'$ is the smallest integer bigger than $f/\omega_0$, $n_0$ is the biggest integer smaller than $N_b/\omega_0$, $\hat{\eta_b}=\mathcal{F}[\eta_b (x,y)]$ is the Fourier transformation of the bottom topography, and $k$ and $l$ are wavenumbers in zonal and meridional directions, respectively \citep{Bell75a,Bell75b}. $\bar{\cdot}$ indicates spatial average. $J_n (\beta)$ is the $n^\mathrm{th}$-order Bessel function, where $\beta\approx U_0 \sqrt{k^2+l^2}/\omega_0$. For Proxima b, the value of $\beta$ varies from the order of 1 to 100. 
\par
\begin{figure}[h!]
\centering
\noindent\includegraphics[width=\textwidth]{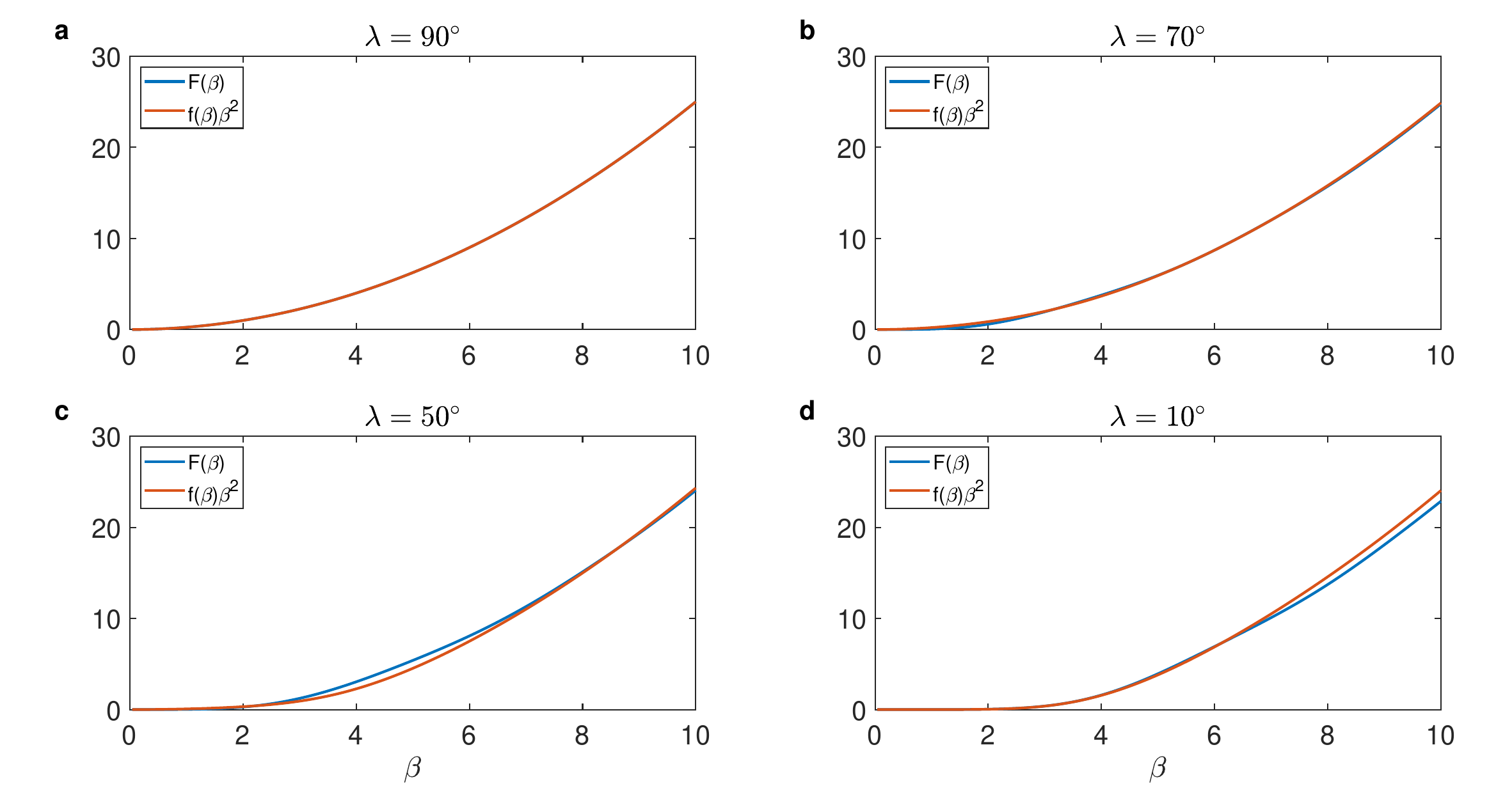}
\caption{Comparison between $F(\beta)$ and $f(\beta)\beta$ defined by Equation~\ref{eq:eq13} at different latitude. $\lambda$ indicates colatitude. The curves are computed with the planetary parameters of Proxima b.}
\label{fig:Fig8}
\end{figure}

Synchronously calculating Equation~(\ref{eq:eq12}) in the ocean tide model is very difficult, so we employ a simpler function to replace the summation in the right hand side of Equation~(\ref{eq:eq12}):
\begin{equation}
    F(\beta)=\sum_{n=n_0'}^{n=n_0}n^2\sqrt{(1-\frac{n^2\omega_0^2}{N_b^2})(1-\frac{f^2}{n^2\omega_0^2})}J_n^2(\beta)\approx \Big(\frac{\sqrt{1-f^2/4\omega_0^2}-1}{1+\beta^4/400}+1\Big)\frac{\beta^2}{4}=f(\beta)\beta^2.
    \label{eq:eq13}
\end{equation}
The functional form of $f(\beta)$ is derived through numerical fitting. The approximation introduced in Equation~(\ref{eq:eq13}) is sufficiently accurate across all latitudes with a deviation within 20\% of the original function (Figure~\ref{fig:Fig8}), provided that the condition $\beta>1$ is met. However, when $\beta\ll1$, Equation~(\ref{eq:eq10}) offers a more precise approximation compared to Equation~(\ref{eq:eq13}). Consequently, for planetary bodies such as Proxima b, TRAPPIST-1e, and GJ 3323b, we adopt Equation~(\ref{eq:eq13}), whereas for Earth, Equation~(\ref{eq:eq10}) is employed to ensure greater accuracy.

\par
\begin{figure}[h!]
\noindent\includegraphics[width=1.0\textwidth]{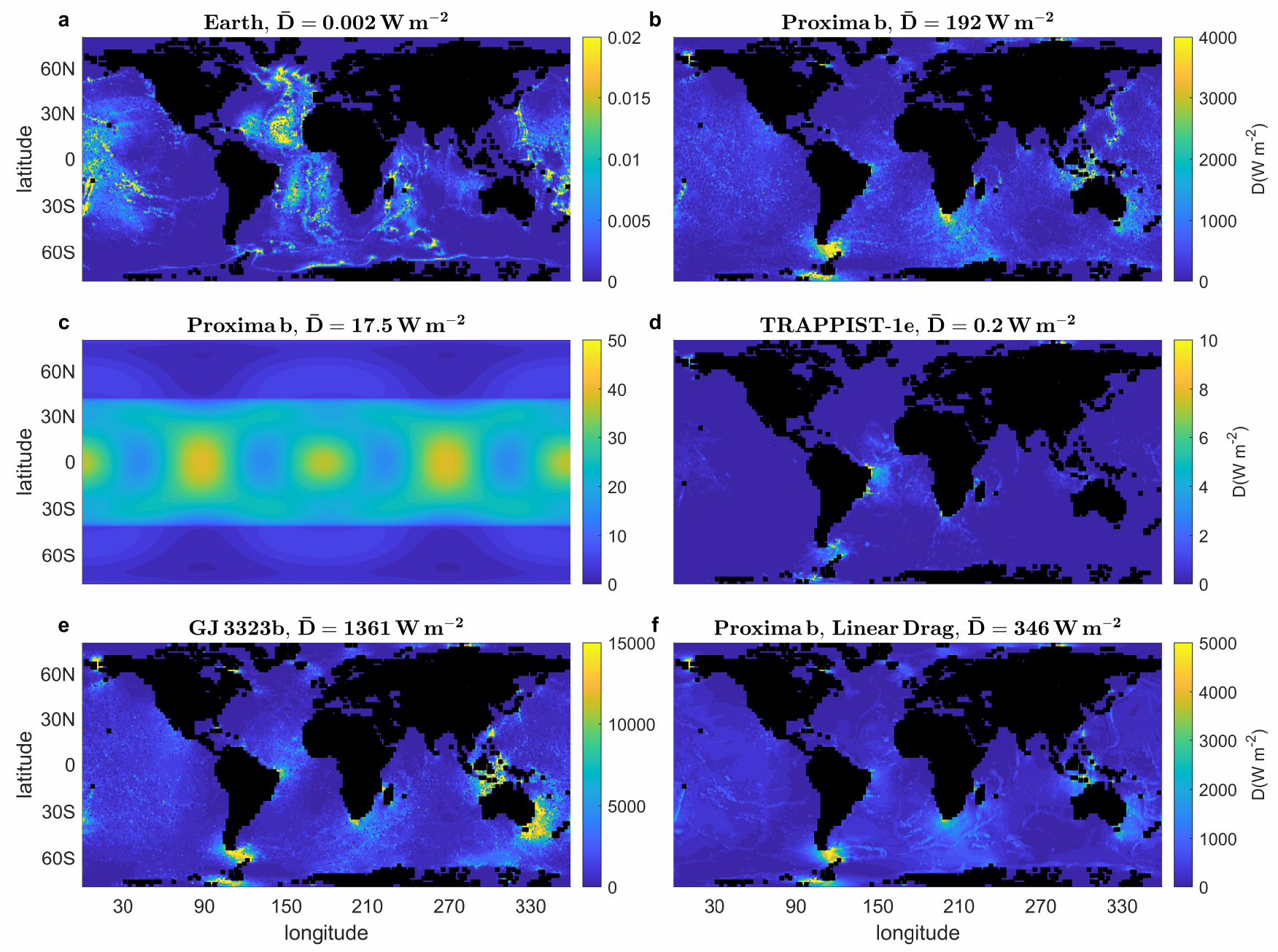}
\caption{Long-term mean tidal dissipation ($D$) on the simulated Earth (a), Proxima b (b, c \& f), TRAPPIST-1e (d), and GJ 3323b (e). $\bar{\cdot}$ denotes spatial average. Panels a, b, d, e \& f are with modern Earth land-ocean configuration, and panel c is for an aqua-planet. Panel f has the same configuration as panel b, but with a simple, linear drag (Equation~(\ref{eq:eq19})). Note that the colorbar ranges differ among the panels. The global-mean value is listed in the center title of each panel. The tidal dissipation is small on Earth and TRAPPIST-1e but large on Proxima b as well as on GJ 3323b.}
\label{fig:Fig5}
\end{figure}

An important consideration arises due to the relatively rapid rotational rates of Proxima b and GJ 3323b compared to their respective orbital rates. Specifically, under these conditions, the coefficient $(1-f^2/(4\omega_0^2))^{0.5}$ becomes imaginary at latitudes exceeding approximately $40^\circ$. The imaginary coefficient can be traced to the first two terms within the summation of Equation~(\ref{eq:eq12}), which correspond to low-frequency internal gravity waves. In high-latitude regions, these waves undergo exponential decay in the vertical direction, implying that they have no contribution to tidal dissipation. To account for this, we impose a constraint wherein $(1-f^2/(4\omega_0^2))^{0.5}$ is set to zero wherever it assumes imaginary values.
According to Equation~(\ref{eq:eq13}), for simulations characterized by large values of $\beta$, the impact of these decaying waves on tidal dissipation is negligible, as illustrated in Figure~\ref{fig:Fig5}b. However, in cases where $\beta$ assumes moderate values--such as in the aqua-planet configuration of Proxima b (Figure~\ref{fig:Fig5}c)--a noticeable reduction in the tidal dissipation rate can be observed around $40^\circ$ latitude. This decline in dissipation highlights the role of decaying waves in modulating energy dissipation processes in planetary atmospheres under varying dynamical conditions.
\par 
Assume $\eta_b$ to be in the form of a Gaussian function:
\begin{equation}
    \hat{\eta_b}=\frac{\eta_bX_0^2}{2}e^{-(k^2+l^2)X_0^2/4},
    \label{eq:eq14}
\end{equation}
Equation~\ref{eq:eq12} can be rewritten as:
\begin{equation}
    \langle P\rangle\approx \frac{\rho_w\omega_0^3N_b\eta_b^2X_0^2}{8\pi^2U_0}\int_0^\infty f(\beta)[e^{-\beta^2\omega_0^2X_0^2/(4U_0^2)}\beta^2]d\beta,
    \label{eq:eq15}
\end{equation}
the part of integrand inside the square bracket peaks at $\sqrt{k^2+l^2}\sim\sqrt 2/X_0$, one may consider $f(\beta)$ as a constant with $\beta=\sqrt2U_0/(\omega_0X_0)$ to further simplify the calculation:
\begin{equation}
    \langle P\rangle\approx \frac{\rho_w\omega_0^3N_b\eta_b^2X_0^2}{8\pi^2U_0}f(\beta)\int_0^\infty e^{-\beta^2\omega_0^2X_0^2/(4U_0^2)}\beta^2d\beta\equiv K'f(\beta)U_0^2,
    \label{eq:eq16}
\end{equation}
where $K'$ is related to seawater density, bottom stratification, and bottom topography:
\begin{equation}
    K'=\frac{\rho_wN_b\eta_b^2}{4\sqrt{\pi}X_0}.
    \label{eq:eq17}
\end{equation}
Finally, based on Equations~(\ref{eq:eq10}), (\ref{eq:eq12}) and (\ref{eq:eq13}), the tidal drag tensor can be calculated as:
\begin{equation}
    \mathbf{C}_\mathrm{tid}\approx \frac{K'f(\beta)u^2}{\rho_w|\boldsymbol{u\cdot\nabla}\eta_b|^2}(\boldsymbol{\nabla}\eta_b)^T\boldsymbol{\nabla}\eta_b.
    \label{eq:eq18}
\end{equation}
The method for calculating the tidal conversion from barotropic tides to internal waves is not unique \citep{Egbert02,Green13}. Another method to parameterize bottom drag is also employed in this work, where the drag coefficient is a scalar:
\begin{equation}    C_\mathrm{tid}\approx\Gamma\frac{(\boldsymbol{\nabla}\eta_b)^2HN_b\bar N}{8\pi^2\omega_0},
    \label{eq:eq19}
\end{equation}
where $\Gamma=50$ is a scaling factor, $N_b$ is buoyancy frequency at ocean bottom, and $\bar N$ is the average buoyancy frequency. Here, $\bar{\cdot}$ indicates average in the vertical direction \citep{Kasting93}. For $N_b$ and $\bar N$, we applied $N=N_0 \exp\{-z/z_0\}$, where $N_0=0.00524\,\mathrm{s^{-1}}$, $z_0=1300\,\mathrm{m}$, $z$ is the seawater depth \citep{Zaron06}. For different climate conditions and for global and annual mean, the values of $N_b$ and $\bar N$ do not change much (see Figure 2 in \citet{Si22}), although there are strong temporal and spatial variations in both $N_b$ and $\bar N$.\par

The tidal dissipation rate per area is calculated through:
\begin{equation}
    D = \rho_w\langle{\boldsymbol{u\cdot}\mathbf{C}_\mathrm{tid}\boldsymbol{\cdot u}}\rangle+\rho_w\langle C_D|\boldsymbol{u}|^3\rangle
    \label{eq:eq20}
\end{equation}
where the first term results from the breakage of internal gravity waves induced by tidal currents, and the second term results from ocean bottom friction (see also Equations~(3) \& (4) in \citet{Green13}). Two different methods were employed to calculate $\mathbf{C}_\mathrm{tid}$, a $2\times2$ drag tensor (by default) and a linear scalar. For the former, Equation~(\ref{eq:eq18}) is used. For the latter, Equation~(\ref{eq:eq19}) is used.

For Earth, the simulated global-mean tidal dissipation $\bar D$ is $\sim0.002\,\mathrm{W\,m^{-2}}$ (Figure~\ref{fig:Fig5}a), consistent with previous simulations \citep{Egbert01,Egbert03}. For Proxima b with modern Earth’s land-ocean configuration, the simulated tidal dissipation is $\sim192\,\mathrm{W\,m^{-2}}$ in global mean and reaches $\sim4000\,\mathrm{W\,m^{-2}}$ in the Cape of Good Hope (Figure~\ref{fig:Fig5}b). In the aqua-planet configuration, the tidal dissipation is about one order of magnitude smaller, $\sim$$17.5\,\mathrm{W\,m^{-2}}$ in global mean (Figure~\ref{fig:Fig5}c).
\par
In the aqua-planet configuration, the spatial distribution of tidal dissipation exhibits a wavenumber-4 pattern in the zonal direction. Specifically, two global maxima are positioned $180^\circ$ apart, while two secondary maxima are symmetrically located at the midpoint between the primary maxima. This distinct spatial inhomogeneity arises as a direct consequence of the 3:2 spin-orbit resonance of Proxima b.
To better understand this pattern, consider the reference frame co-rotating with Proxima b, wherein the host star appears to orbit the planet at an average angular velocity 0.5 times the orbital rate of the planet. Within a single orbital cycle, the star thus experiences two periastron and two apastron passages. This configuration implies that Proxima b locates at the center of the elliptical trajectory of the star, with the perihelia positioned at the endpoints of the minor axis and the aphelia at the endpoints of the major axis of the elliptical orbit.
As a result, the amplitude of the tidal potential, as defined in Section 2.3, exhibits a strong longitudinal dependence on Proxima b. More precisely, it attains local maxima at the four distinct longitudinal locations corresponding to the instances when the star reaches its periastron or apastron (see Figure~\ref{fig:Fig1}c). This longitudinal inhomogeneity in tidal potential amplitude ultimately governs the spatial distribution of tidal dissipation, thereby giving rise to the wavenumber-4 pattern observed in Figure~\ref{fig:Fig5}c.
\par

In the simulation of GJ 3323b, the global-mean tidal dissipation is $1361\,\mathrm{W\,m^{-2}}$ under modern Earth’s land-ocean configuration (Figure~\ref{fig:Fig5}e) and $58\,\mathrm{W\,m^{-2}}$ in aqua-planet configuration (not shown). These are stronger than those for Proxima b, because GJ 3323b has higher tidal potential and larger tidal currents. In the simulation of TRAPPIST-1e, the global-mean tidal dissipation is only $0.2\,\mathrm{W\,m^{-2}}$ (Figure~\ref{fig:Fig5}d) due to the small eccentricity (0.005).
\par

By default, tidal drag coefficient is a $2\times2$ tensor and it is a complex function of current speed, tidal frequency, ocean stratification, and seafloor roughness (Equation~(\ref{eq:eq18})). If a simplified linear drag (Equation~(\ref{eq:eq19})) is employed, the estimated global-mean tidal dissipation rate is $346\,\mathrm{W\,m^{-2}}$ for Proxima b with modern Earth’s land-ocean configuration (Figure~\ref{fig:Fig5}f), which is 80\% higher than that from the default method. 
\par
Note that even for Earth, present understanding of tidal dissipation processes is poor, so the uncertainty is unavoidable \citep{Green13}. Nonetheless, our results suggest that the tidal dissipation rate is in the order of $100\,\mathrm{W\,m^{-2}}$ for Proxima b with continents if assuming the planet has a high eccentricity, which is comparable to the global-mean stellar flux, $\sim$$225\,\mathrm{W\,m^{-2}}$. Our results are consistent with that of \citet{Barnes13}, who also showed that tidal dissipation on planets with high eccentricities may reach a high-enough level to trigger a runaway greenhouse. For planets with small eccentricities, the tidal dissipation is weak.\par

\subsection{Orbital Evolution}
Previous studies have demonstrated that oceanic tidal dissipation plays a fundamental role in governing the long-term orbital evolution of the Earth-Moon system \citep{Webb80,Webb82,Blackledge20,Motoyama20,Tyler21,Daher21}. This well-established influence naturally raises the question of whether a similar mechanism may also operate within exoplanetary systems.
To estimate the impact of tidal forces on planetary orbits, two commonly employed approaches are the constant phase lag (CPL) model and the constant time lag (CTL) model \citep{Ferraz-Mello08,Leconte10}. These methods rely on either an assumed tidal quality factor ($Q$) or an assumed tidal time lag ($\tau$) to compute the tidal dissipation rate ($\bar D$), as well as to track the resulting evolution of the planetary orbit and rotational state over time. Both approaches have been extensively documented and applied in previous studies, particularly in \citet{Barnes13} and \citet{Barnes17}.
\par
\begin{figure}[h!]
\centering
\noindent\includegraphics[width=0.7\textwidth]{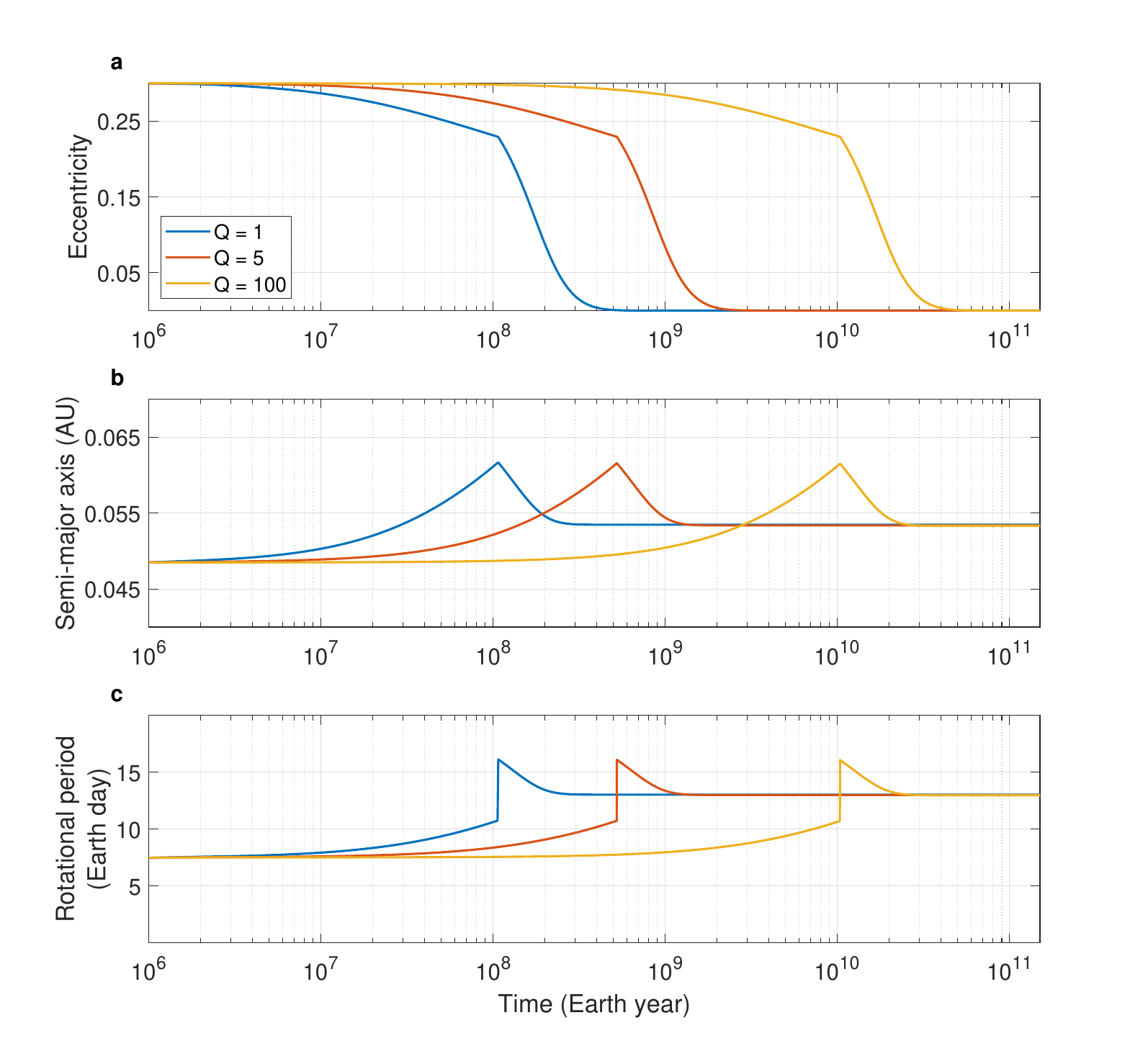}
\caption{Orbit evolution of Proxima b under three different values of tidal quality factor ($Q$): 100 (yellow), 5 (red), and 1 (blue). (a) Orbital eccentricity, (b) semi-major axis, and (c) rotation period. The initial eccentricity is 0.30, semi-major axis is 0.0485 $\mathrm{AU}$, and rotation period is 7.46 Earth days. The kinks when the eccentricity is close to 0.23 result from the fact that only the first four Fourier components of the tidal torque are included in the CPL model (see \citet{Barnes13} and \citet{Barnes17}). Increasing tidal dissipation rate (corresponding to a smaller $Q$) strongly accelerates the orbital evolution.}
\label{fig:Fig7}
\end{figure}

In this study, we employ the CPL model with the software package EQTIDE \citep{Barnes17}. Based on angular momentum conservation, the model has six differential equations governing the variation of six independent variables: planetary eccentricity, semi-major axis between star and planet, rotational rate and obliquity of the star, and rotational rate and obliquity of the planet (see Equations (E1)-(E4) in \citet{Barnes13}). The results are shown in Figure~\ref{fig:Fig7}.
One key parameter of the CPL model is the value of $Q$. In this study, the value of $\bar D$ can be obtained from the ocean tide model above, so we could use an inverse method to find $Q$. According to \citet{Barnes13}, the dissipation rate in rotational and orbital energy caused by the tides are:
\begin{equation}
    \dot E_\mathrm{rot} = \frac{3G^2k_2M^2(M+m_p)a^5}{8r^9\bar \omega Q}\Big[4\epsilon_0+e^2\Big(-20\epsilon_0+\frac{147}{2}\epsilon_1+\frac{1}{2}\epsilon_2-3\epsilon_5\Big)\Big],
    \label{eq:eq21}
\end{equation}
\begin{equation}
    \dot E_\mathrm{orb} = -\frac{3G^2k_2M^2(M+m_p)a^5}{8r^9\bar \omega^2Q}\Omega\Big[4\epsilon_0+e^2\Big(-20\epsilon_0+49\epsilon_1+\epsilon_2\Big)\Big],
    \label{eq:eq22}
\end{equation}
where $G$ is the gravitational constant, $k_2$ is the Love number of the planet, $M$ is the mass of the star, $m_p$ is the mass of the planet, $a$ is the mean radius of the planet, $r$ is the semi-major axis of the orbit, $\bar\omega$ is mean orbital rate, $\Omega$ is the rotational rate of the planet, $e$ is orbital eccentricity, and $\epsilon_0$, $\epsilon_1$, $\epsilon_3$ and $\epsilon_5$ are step functions ($\pm1$ or 0) indicating the signs of the phase lags (see Equation (E7) in \citet{Barnes13}). The sum of $\dot E_\mathrm{rot}$ and $\dot E_\mathrm{orb}$ per square meter is approximately equal to the value of $\bar D$.\par

For Earth, $\Omega\gg\bar\omega$, we have $\epsilon_0=1$, $\epsilon_1=1$, $\epsilon_2=1$, and $\epsilon_5=1$. With planetary parameters shown in Table~\ref{tab:Tab1} and with $\bar D=0.002\,\mathrm{W\,m^{-2}}$, we obtain:
\begin{equation}
    Q = \frac{3G^2k_2M^2(M+m_p)a^5}{8r^9\bar\omega\cdot4\pi a^2\bar D}(1456+10899e^2)\approx 10,
    \label{eq:eq23}
\end{equation}
which is close to the estimated value for Earth \citep{Goldreich66}.\par

For Proxima b, $\Omega=1.5\bar\omega$, we have $\epsilon_0=1$, $\epsilon_1=0$, $\epsilon_2=1$, and $\epsilon_5=1$. For control experiment of Proxima b with modern Earth's land-ocean configuration, $\bar D=192\,\mathrm{W\,m^{-2}}$, we obtain:
\begin{equation}
    Q = \frac{3G^2k_2M^2(M+m_p)a^5}{8r^9\bar\omega\cdot4\pi a^2\bar D}(2-6e^2)\approx 5.
    \label{eq:eq24}
\end{equation}
Note that, in order to keep $Q$ positive, Equation~(\ref{eq:eq24}) cannot be applied to an eccentricity higher than 0.577.

The estimated $Q$ of our experiments covers a wide range from 0.2 to 2000, strongly depending on land-ocean configuration, ocean depth, planetary rotation rate, orbital eccentricity, and other factors (Table~\ref{tab:Tab3}). In principle, $Q$ should exceed 1, as it is defined as the ratio of the energy stored within the tidal distortion to the energy dissipated over one tidal period \citep{Goldreich66}. This definition underlies Equations~(\ref{eq:eq21}) and~(\ref{eq:eq22}). However, the tidal potential energy in our formulation is derived from equilibrium tide theory, which assumes that the ocean surface closely follows the equilibrium tidal shape (i.e., an equipotential surface of the tidal forcing). While this assumption holds for most of our experiments, it may break down under conditions of tidal resonance (marked by note $c$ in Table~\ref{tab:Tab3}), potentially resulting in values of $Q < 1$.

The resultant orbital evolution under three different values of $Q$, 100 for a dry rocky planet, 5 for a wet planet with ocean tides (Earth’s $Q$ is $\sim12$, mainly from ocean tides \citep{Goldreich66}), and 1 for a wet planet with extremely strong ocean tides, are examined (Figure~\ref{fig:Fig7}). Ocean tides accelerate orbital evolution by $\sim20$ times for $Q$ = 5 and by $\sim100$ times for $Q$ = 1, compared with that for $Q$ = 100. This implies that ocean tides can significantly shorten the time scale for the planet to enter a 1:1 tide-locked state, from $10^{10}$ to $10^9$ or $10^8$ Earth years.\par

\section{Conclusions and Discussions} \label{sec:conclusion}
In this work, we find that ocean tides on asynchronously rotating planets orbiting low-mass stars can reach $\mathcal{O}(1000)\,\mathrm{m}$ in height, tidal currents can reach $\mathcal{O}(10)\,\mathrm{m\,s^{-1}}$, if the orbital eccentricities are large. Under this condition, it is likely that all continents above the sea level would be easily eroded away by ocean tides, leaving no continent such that the planet would be more like an aqua-planet if the planet still has oceans \citep{Dole64}. These tides may dramatically influence orbital evolution, climate, habitability, and observational features. The strong ocean tides can induce large tidal energy dissipation (see Figure~\ref{fig:Fig5}) and can also enhance deep ocean mixing, oceanic overturning circulation and thereby equator-to-pole ocean heat transport \citep{Si22,Barnes23}. For asynchronously rotating planets with small eccentricities, the ocean tides would be much weaker but is still non-negligible and comparable to that on Earth.\par

One weakness in our numerical simulations is that the model we employed has only one layer in the vertical direction, so ocean's real stratification is not captured. Another potential issue associated to our model configuration is that, it is likely for tidal flow to penetrate through the coastline over a large distance, given the high tide (100-1000 m) on the exoplanets considered in this work. Future work is required to more accurately simulate the response of the whole climate system to the strong oceanic tides as well as atmospheric tides \citep{Cunha14,Leconte15}. 

Finally, here we focus on the two-body system (one star and one planet), and the possible effects of additional planets or moons are not considered. Two possible consequences of multi-body interactions are asynchronous rotation \citep{Chen21} and high obliquity \citep{Millholland24}. Asynchronous rotation could enhance eccentricity tide, while high obliquity could drive strong obliquity tide. In contrast to eccentricity tide--where tidal bulge moves back and forward along equator--obliquity tide features meridional oscillation of tidal bulge. Future work is required for multi-body interactions.


\begin{acknowledgments}
We thank MIT team for the release of MITgcm. We are grateful for the help from Daniel Koll, Xing Wei, Feng He, Mingyu Yan, Yidongfang Si, Yaoxuan Zeng and Nan Zhang. This work was partially motivated by the movie directed by Christopher Nolan in 2014, “Interstellar”. J.Y. is supported by National Science Funding of China under grant 42441812 and by National Key Research and Development Program of China under grant 2024YFF0807903.\par
The oceanic tide experiments in this work are run on a modified version of the software MITgcm available at \url{http://mitgcm.org}, the modified codes are available at \url{https://doi.org/10.5281/zenodo.15476076}. Simulation outputs used to create the figures are available at \url{https://doi.org/10.5281/zenodo.15467971}. The orbital evolution experiments in this work are run on the software package EQTIDE available at \url{https://github.com/RoryBarnes/EqTide}.
\end{acknowledgments}




\bibliography{Main_text_Jiaru}{}
\bibliographystyle{aasjournal}



\end{document}